# A Pressure Ulcer Care System For Remote Medical Assistance: Residual U-Net with an Attention Model Based for Wound Area Segmentation

**Jinyeong Chae[1], Ki Yong Hong[2], Jihie Kim[1]**

[1]Department of Artificial intelligence, University of Dongguk

[2]Department of Plastic and Reconstructive Surgery, Dongguk University Ilsan Hospital

jiny419@dgu.ac.kr, pskyhong@gmail.com, jihie.kim@dgu.edu

## Abstract

Increasing numbers of patients with disabilities or elderly people with mobility issues often suffer from a pressure ulcer. The affected areas need regular checks, but they have a difficulty in accessing a hospital. Some remote diagnosis systems are being used for them, but there are limitations in checking a patient's status regularly. In this paper, we present a remote medical assistant that can help pressure ulcer management with image processing techniques. The proposed system includes a mobile application with a deep learning model for wound segmentation and analysis. As there are not enough data to train the deep learning model, we make use of a pre-trained model from a relevant domain and data augmentation that is appropriate for this task. First of all, an image pre-processing method using bilinear interpolation is used to resize images and normalize the images. Second, for data augmentation, we use rotation, reflection, and a watershed algorithm. Third, we use a pre-trained deep learning model generated from skin wound images similar to pressure ulcer images. Finally, we added an attention module that can provide hints on the pressure ulcer image features. The resulting model provides an accuracy of 99.0%, an intersection over union (IoU) of 99.99%, and a dice similarity coefficient (DSC) of 93.4% for pressure ulcer segmentation, which is better than existing results.

## Introduction

As technology advances, telehealth approaches are increasingly attempted as a medical assistant. The system can potentially allow patients to check their medical conditions at home rather than visiting a hospital physically. Such a system is particularly imperative for patients with disabilities or mobility issues. As some patients with disabilities are paralyzed or live in a bed, some part of their body is constantly pressured. These skin wounds are called pressure ulcer (PU). The existence, size, and depth of the PU are important for these patients because the diagnosis method varies depending on the information. As these patients have a difficulty in accessing a hospital but need continuous monitoring and management, a remote management system is very needed. In this study, we propose a remote medical assistant that allows patients or their assistants to continuously manage the wound by capturing the image using their mobile phones. Our system architecture is shown in Fig 1. A deep learning model analyzes and segments the images collected by the patients, and the segmented wound images are visualized. The visualized images help a doctor assess the wound area and suggest how to treat the wound or when to visit the hospital. We expect that by using such automated analyses of the wound area, a remote assessment and treatment by a doctor can be done more efficiently.

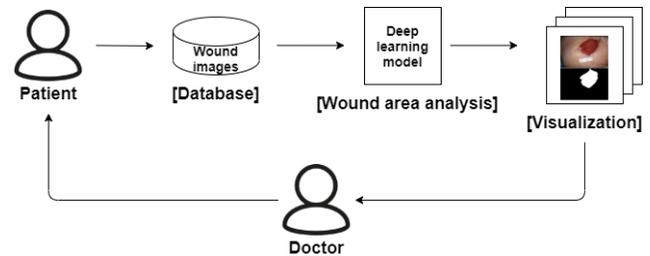

Fig 1. A remote medical assistant for pressure ulcer care

Existing remote diagnosis technology (Wang et al. 2018) makes use of a questionnaire form that asks about the patient's wound condition. The existence, depth, and size of a wound are evaluated based on the patient's judgment. Appropriate treatment methods are provided according to the answers of the patient. However, in such systems, patients manually fill in the information and the system doesn't analyze the wound image. This study proposes a remote management system that uses the patient's wound images and applies a deep learning model to segment the wound area. The segmented wound area is crucial as it can be used to measure the size of the wound and to assess the progress. In addition, visualizing the segmented wound areas can assist a doctor to evaluate the wound's status.

Semantic image segmentation is a high-level problem for classifying photos and understanding the overall photograph` scenes (Liu et al. 2018). The purpose of semantic image segmentation is to classify all the pixels into a specified number of classes. It is also called pixel-level classification as a prediction for all the pixels. Image segmentation is used in various fields, such as land cover classification and road signs detection. It is also used in medical fields such as medical device detection during surgery, brain tumors, and bedsore detection. However, due to a lack of relevant patient data, it is necessary to study efficient image segmentation and image processing techniques with limited data. Our study focuses on image segmentation for medical images where data is scarce, and this paper presents the results in localizing bedsore or PU. Several deep learning approaches have been proposed for segmenting or classifying wound images. For example, (Wang et al. 2015) studied the end-to-end method based on convolutional neural networks (CNN) using 2,700 wound images. The Intersection of Union of the study was 47.3%. (Pholberdee et al. 2018) also makes use of CNN model but performs data augmentation using color variation given fewer image datasets than (Wang et al. 2015). Their method achieved up to 53% IoU. (Goyal et al. 2017) proposed automated diabetic foot ulcer and its surrounding skin segmentation by using FCN(Fully Connected Networks). In performance measures, FCN-16s was the best model for achieving DSC of 79.4% for the ulcer and surrounding skin region. However, the accuracy is low when segmenting wounds with irregular borders because the classified area is smooth contours. (Garcia-Zapirain et al. 2018) presented a classification framework with multiple pathways for PU images based on 3D CNN. The network used to train 193 images and showed 92% of DSC, 95% of AUC. (Khalil et al. 2019) proposed various feature extraction methods based on color, texture, and it classified four types of wound tissues. This study showed an average accuracy of 96%.

In our work, for an automatic learning system, we use Residual U-Net with an attention module rather than a simple CNN model with a limited dataset. We first perform image pre-processing techniques using a medical expert's visual method, observing the PU's size and color. We also apply data augmentation techniques that are appropriate for medical images, such as rotation, reflection, and a watershed algorithm. In addition, we make use of a pre-training method using similar wound images. For generating segmentation models, we add an attention module to help the model focus on relevant visual features. The resulting model provides an accuracy of 99.0%, IoU of 99.9% and DSC of 93.4% for pressure ulcer segmentation, which is better than the existing approaches. This work's contribution is two folds: First, we propose a method for combining image pre-processing, data augmentation, and an attention model for handling insufficient image data for pressure ulcer segmentation. Second, we applied them to PU image datasets and show that it is more effective than other current approaches. The following section describes related works. The description of the data used in experiments is provided in the data section. We also discuss our method' details in the method section, while experiments and results are given in the results section. Finally, we present the conclusions and future works in the last section.

## Related works

Image segmentation refers to dividing an image into pixels and identifying a region of interest in computer vision. Image segmentation methods include fully convolutional networks (FCN) (Long et al. 2015) and U-Net (Ronneberger et al. 2015). These methods can also be useful for medical image analyses. The quality of the analysis can be improved by conveying important information about the lesion's size and shape. Much research has been done on the segmentation of lesions. Segmentation can identify the scope and the shape of the region of interest and help label the area (Hesam-Hesamian et al. 2019). The most important characteristic of U-Net is the connection between the same resolutions layers and it provides the high resolutions features to the deconvolutional layer (Hesam-Hesamian et al. 2019). Because of these characteristics, the U-Net structure is more suitable for medical image segmentation than (Lee et al. 2018). (He et al. 2016) proposed an improved model to solve the gradient vanishing problem that occurs when the layer becomes deeper. The authors added a skip connection between layers to deliver the gradient well. In (Huang et al. 2017), the authors proposed densely connected convolutional networks (DenseNet), which is inspired by the residual neural network (ResNet) idea. It applied a skip connection to all the layers of the entire network. Our work is inspired by the feature extraction in which the decoder concatenates the encoder's feature. We used it for down-sampling and up-sampling. The difference from the previous study is to design a model that segments to focus only on the wound region by adding skip connection to convolution block and an attention module to the decoder, as shown in Fig 2.

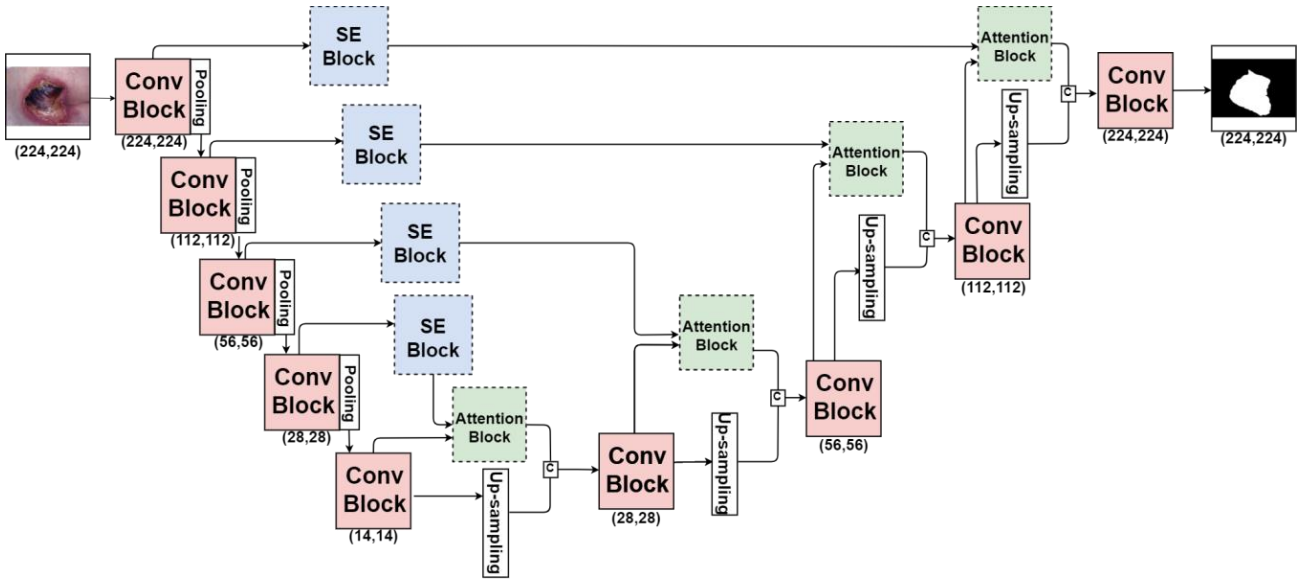

Fig 2. Residual U-Net with an Attention module

For automated PU analyses, several image processing approaches have been applied, including traditional computer vision technologies (David et al. 2017) (Elmogy et al. 2018). (David et al. 2017) performed PU image segmentation based on contrast changes computed using synthetic frequencies. The authors applied various morphological operations such as erosion and dilation for the decomposition of the images. From the original image, once the lines were extracted from the entire image, the blurring boundary is cleaned using a dilation technique. Then, by applying erosion, unnecessary pixels outside the wound were eliminated and converted into a background image. As a test image, 51 sheets were used and showed an average correlation of 0.89. In this study, inspired by the boundary processing method (David et al. 2017), we use a marker-based watershed algorithm that augments data by distinguishing between a wound region and a general skin area (Lalitha et al. 2016)(Bai and Urtasun 2017)(Fabijanska 2012). In (Mukherjee et al. 2014), a chronic wound image was segmented using fuzzy divergence based on the thresholding technique. During the pre-processing of the image, the [red, green, blue] (RGB) image was converted to [hue, saturation, intensity] (HIS). The authors said that by altering the RGB image to the HIS values, they could more accurately extract the wound's boundary. The resulting segmentation accuracy was 87.61%. In our work, PU pictures are taken with different devices (personal cell phones), and the image's brightness varies. Therefore, we used original RGB images rather than changing the color space HSI. (Goyal et al. 2017) used fully convolutional neural networks using 705 images and achieved 90% DSC. They implemented a two-tier transfer learning method but used an image dataset from an irrelevant domain. (Garcia-Zapirain et al. 2018) presented 3D CNN with multiple pathways using 193 images and showed 92% DSC. They presented various features along with different modalities. However, the authors did not discuss the effective method given the limited dataset. (Khalil et al. 2019) also proposed various feature extraction based on color and texture. The authors classified four types of tissues with an accuracy of 96%. But this method is not end-to-end learning for an automatic system. Furthermore, they did not use an attention method to improve performance.

There are also a lot of image augmentation techniques due to a problem of a lack of image data. In (Shorten et al. 2019), the authors used flipping, cropping, rotation, and noise injection. There is also a study of segmenting a wound image using image data augmentation. In (Shenoy et al. 2018), the number of images was increased using five methods of rotation, shifting, zooming, shearing, and flipping. In (Pholberdee et al. 2018), the color variation method was applied to the wound image to increase training data size three times. In (Zhang et al. 2018), the authors augmented the wound image using DCGANs as a deep learning model approach. In our study, since we handle medical images, appropriate data augmentation techniques such as flipping and rotation are used. We applied reflection based on the x-axis and y-axis and rotated at an angle between -90 and 90 on the image. Furthermore, we applied a watershed algorithm to augment the data by making the boundary between the wound and the normal skin clear.

U-Net was recently proposed in (Ronneberger et al. 2015) to segment biomedical images and is currently being actively used as a state-of-the-art technique. In this paper, we created a baseline model with U-Net for the segmentation of PU images. We also apply an attention mechanism in the decoder. This mechanism mitigates the problem of information loss when compressing information into a fixed-sized vector. The attention mechanism refers to the encoder information whenever the decoder predicts the output information. In (Oktay et al. 2018), the authors used an attention mechanism to a decoder. In this study, we apply a similar attention mechanism by concatenating an up-sampling layer, and attention modules are used to extract critical features for PU images. However, unlike the attention method that existed in (Oktay et al. 2018), we used an attention module that can extract both the channel and spatial features of the image to focus on the wound region, not just a simple attention block.

## Data

In this study, we used three types of datasets. First, we used PU images collected by Dongguk University Ilsan Hospital. The PU image dataset consists of a total of 101 RGB images. Second, Medetec Wound Database (MWD) was used (Thomas 2019). The MWD is an image dataset with various other wound types such as venous leg ulcer, arterial leg ulcer, malignant wound, and surgical wound infection. It consists of 264 images with a resolution of 1024x731. Third, 1109 wound images collected by Azh wound care center (AZH) are also used. A pre-training was performed with the wound images of the MWD and the AZH. In addition, we performed training and testing on the PU images.

## Method

**Image pre-processing**

All the original images are transformed with 224 x 224 size for training since the actual images were different in size. In addition, we used a bilinear interpolation method to resize images. Also, the pixel brightness was adjusted during the process of resizing the image. For the pixels lumped together, the anti-aliasing technique was used to find the wound's boundary (Trusov et al. 2020). After image pre-processing, data augmentation techniques such as image rotation and reflection are used to solve data shortage. We used a random value between -90 and 90 to increase images for the rotation. We also used a watershed algorithm with a high threshold for additional data augmentation techniques (Fabijanska 2012).

**Proposed model**

In this paper, we propose a Residual U-Net model with an attention module in Fig 2. The images resized become the model's input, and the input data are passed through the convolutional block and attention module. The output is an image segmented for the PU region. Our model is divided up to encoder and decoder based on the center of Fig 2. The encoder in the model is the process of down-sampling the image. The down-sampling has a convolution block of 2-D to extract features. And the convolution block shown in Fig 3 includes a skip connection to reflect the original features. The decoder has an attention module and a convolutional block of 2-D. In the decoder, the SE block and attention block are used to extract the image's channels and spatial features. In addition, when up-sampling is implemented in the decoder, the encoder's attention module was concatenated to the decoder for better up-sampling, as shown in Fig 2. Therefore, our model architecture can focus on the important area of the image and segment of the PU region.

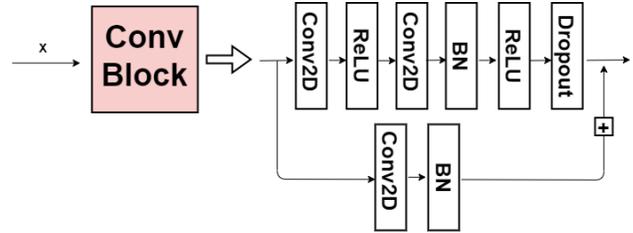

Fig 3. Convolution block

**Pre-training**

Based on the fact that weights learned from similar images can improve the network's performance, this study proposes a pre-training method. Since the number of training data required for the segmentation of the PU area is insufficient, our model is pre-trained with similar wound images. The weights learned from the wound image are used to segment the PU image. In this study, we used PU images after pre-training our model using AZH and MWD images.

**Attention module**

An attention module consists of two blocks: a SE block and an attention block. As shown in Fig 4., The SE block extracts channel information through a global average pooling layer and batch normalization layer (BN) and it calculates each channel's importance through a fully connected layer. Thus, the SE block extracts channel information reflected to which channel affects a lot. The attention block extracts spatial information. The block consists of two inputs. One is channel information extracted from the SE block, and the other is the convolution block. As shown in Fig 5., the attention block extracts spatial features reflected the channel features through the convolutional layer. Therefore, the attention module considers both the channel and spatial features of the previous layer. In addi-tion, the features are reflected in the up-sampling process by concatenation.

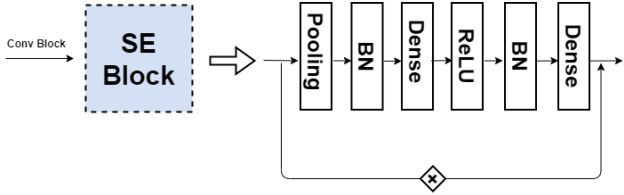

Fig 4. Squeeze-Excitation block

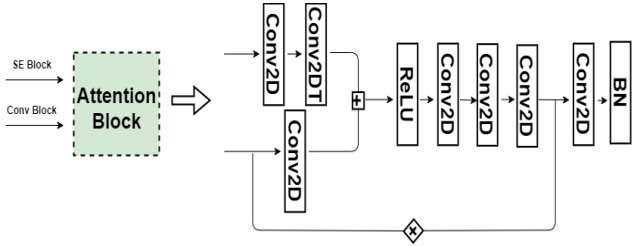

Fig 5. Attention block

## Results

The pixel-level evaluation metrics are Accuracy (Acc), Intersection over Union (IoU), and Dice similarity coefficient (DSC). The following equations (1), (2), and (3) are metrics.
The experiments were carried out on the PU dataset split into 70% training, 10% validation, and 20% testing.

$$IoU = Area\ of\ Overlap / Area\ of\ Union \quad (1)$$
$$DSC = 2 * Area\ of\ Overlap / Total\ pixels\ combined \quad (2)$$
$$Acc = (TP + TN) / (TP + TN + FP + FN) \quad (3)$$

Table 1. Comparison of results with pre-training

| Experiments | Acc(%) | IoU(%) | DSC(%) |
|---|---|---|---|
| w/ pre-training | **99.0** | **99.9** | **93.4** |
| w/o pre-training | 98.5 | 99.9 | 62.0 |

Table 1. shows a comparison of results with pre-training. The results with the pre-training method show 31.4% better performance than without pre-training in DSC.

Table 2. Comparison of results with an attention module

| Experiments | Acc(%) | IoU(%) | DSC(%) |
|---|---|---|---|
| w/ attention | **99.0** | **99.9** | **93.4** |
| w/o attention | 98.9 | 99.9 | 93.3 |

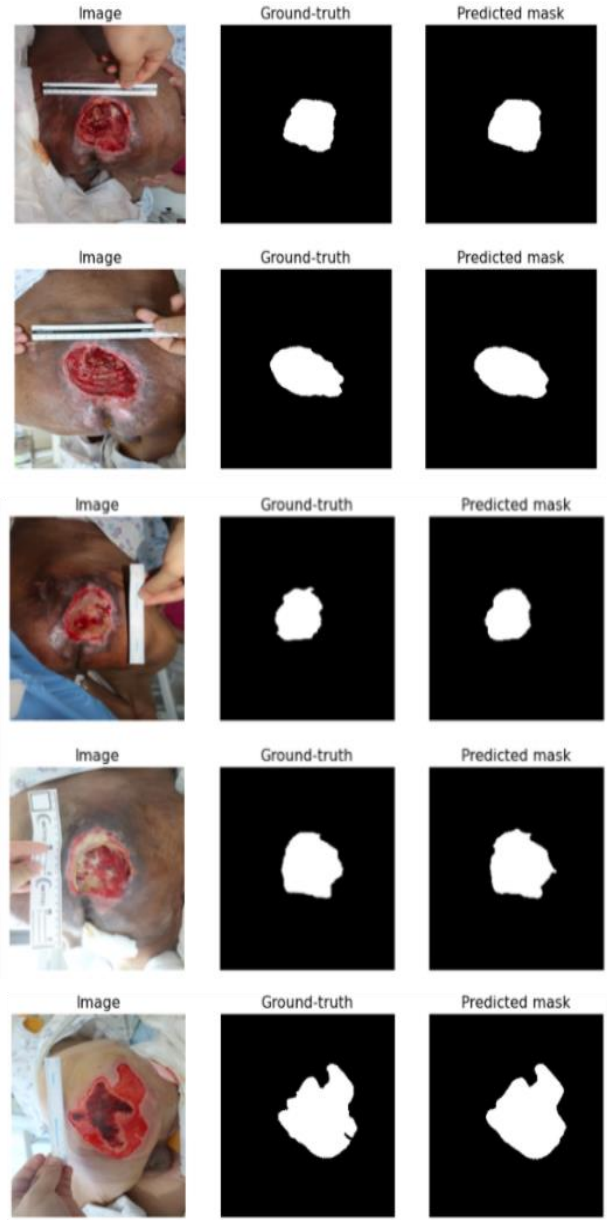

Fig 6. Segmentation results

Table 3. Results compared with state-of-the-art methods

| Experiments | Acc(%) | IoU(%) | DSC(%) |
|---|---|---|---|
| Wang et al. 2015 | 95.0 | 47.3 | - |
| Pholberdee et al. 2018 | 99.0 | 99.7 | 75.3 |
| Goyal et al. 2017 | - | - | 89.9 |
| Garcia-Zapirain et al. 2018 | - | - | 92.0 |
| Khalil et al. 2019 | 96.0 | - | - |
| Residual U-Net + Attention | **99.0** | **99.9** | **93.4** |

We also show an attention module performance in Table 2. After applying the module, it shows 0.1%, 1.1% better performance in Acc and DSC than before. The segmented PU regions are shown in Fig 6., and the performance results are presented with previous studies in Table 3. As shown in Table 3, our model outperforms other state-of-the-art techniques. Our approach that combines image pre-processing and data augmentation showed a significant 3% of Acc, 0.2% of IoU, and 1.4% of DSC better performance than the previous study.

## Conclusion and Future works

In this study, we proposed remote medical assistance system using an image segmentation deep learning model for PU images. We also presented an approach for making the most out of the available dataset when the given image dataset is limited. We combined an image pre-processing, data augmentation, and an attention module. By applying this method to PU image segmentation, we achieved performance improvement over existing approaches. We expect that the approach can be used to similar image processing problems when the given image dataset is limited. As future works, we plan to study a model that can further improve the pressure ulcer analysis by using temporal changes. We are also interested in researching status changes of the patient's wound by estimating the size and the depth of the wound from a limited image dataset.

## Acknowledgement

This research was supported by the MSIT(Ministry of Science and ICT), Korea, under the ITRC(Information Technology Research Center) support program(IITP-2020-2020-0-01789) supervised by the IITP(Institute of Information & Communications Technology Planning & Evaluation)

## References

[Wang et al. 2018] Wang S. et al. 2018. A New Smart Mobile System for Chronic Wound Care Management. in *IEEE Access*. 6. 52355-52365.

[Liu et al. 2018] Liu, X.; Deng, Z.; and Yang, Y. 2018. Recent progress in semantic image segmentation. Artificial Intelligence Review, 1- 18.

[Wang et al. 2015] Wang, C.; Yan, X.; Smith, M.; Kochhar, K.; Rubin, M.; Warren, S.-M.; Wrobel, J.; Lee, H.; 2015. A unified framework for automatic wound segmentation and analysis with deep convolutional neural networks. 2015 37th Annual International Conference of the IEEE Engineering in Medicine and Biology Society (EMBC), 2415-2418.

[Pholberdee et al. 2018] Pholberdee, N.; Pathompatai, C.; and Taeprasartsit, P.; 2018. Study of Chronic Wound Image Segmentation: Impact of Tissue Type and Color Data Augmentation. 15th International Joint Conference on Computer Science and Software Engineering (JCSSE), 1-6.

[Goyal et al. 2017] Goyal, M.; Yap, M.-H.; Reeves N.-D.; Rajbhandari, S.; Spragg, J.; 2017. Fully convolutional networks for diabetic foot ulcer segmentation. IEEE International Conference on Systems, Man, and Cybernetics (SMC), 618-623.

[Garcia-Zapirain et al. 2018] Garcia-Zapirain, B.; Elmogy, M.; El-Baz, S.-A.; and Elmaghraby A.-S.; 2018. Classification of pressure ulcer tissues with 3D convolutional neural network. Medical & Biological Engineering & Computing, 1-14.

[Khalil et al. 2019] Khalil, A.; Elmogy, M.; Ghazal, M.; Burns, C.; and El-Baz, A.; 2019. Chronic Wound Healing Assessment System Based on Different Features Modalities and Non-Negative Matrix Factorization (NMF) Feature Reduction. *IEEE Access*. 7. 80110-80121.

[Long et al. 2015] Long, J.; Shelhamer, E.; and Darrell, T.; 2015. Fully convolutional networks for semantic segmentation. *IEEE Conference on Computer Vision and Pattern Recognition (CVPR)*. 3431-3440.

[Ronneberger et al. 2015] Ronnberger, O.; Fischer, P.; and Brox, T.; 2015. U-Net: Convolutional Networks for Biomedical Image Segmentation. International Conference on Medical Image Computing and Computer Assisted Intervention(MICCAI). 234-241.

[Hesam-Hesamian etal. 2019] Hesamian-Hesam, M.; Jia, W.; He, X.; and Kennedy, P.; 2019. Deep Learning Techniques for Medical Image Segmentation: Achievements and Challenges. Journal of Digital Imaging 32. 582-596.

[Lee et al. 2018] Lee, Y.-C.; Jung, S.-H.; and Won, H.-H.; 2018. WonDerM: Skin Lesion Classification with Fine-tuned Neural Networks. ISIC 2018.

[He et al. 2016] He, K.; Zhang, X.; Ren, S.; and Sun, J.; 2016. Deep Residual Learning for Image Recognition. IEEE Conference on Computer Vision and Pattern Recognition (CVPR). 770-778.

[Huang et al. 2017] Huang, G.; Liu, Z.; Van Der Maaten, L.; and Weinberger, K.-Q.; 2017. Densely Connected Convolutional Networks. IEEE Conference on Computer Vision and Pattern Recognition (CVPR). 2261-2269.

[David et al. 2017] David, O.-P.; Sierra-Sosa, D.; and Garcia-Zapirain, B.; 2017. Pressure ulcer image segmentation technique through synthetic frequencies generation and contrast variation using toroidal geometry. BioMed Eng Online16. 4.

[Elmogy et al. 2018] Elmogy, M.; García-Zapirain, B.; Burns, C.; Elmaghraby, A.; and Ei-Baz, A.; 2018. Tissues Classification for Pressure Ulcer Images Based on 3D Convolutional Neural Network. IEEE International Conference on Image Processing (ICIP). 3139-3143.

[Lalitha et al. 2016] Lalitha, K.-V.; Amrutha, R.; Michahial, S.; and Dr, M.; 2016. Implementation of Watershed Segmentation. IJARCCE 5. 196-199

[Bai and Urtasun 2017] Bai, M.; and Urtasun, R.; 2017. Deep Watershed Transform for Instance Segmentation. IEEE Conference on Computer Vision and Pattern Recognition (CVPR). 2858-2866

[Fabijanska 2012] Fabijanska, A.; 2012. Normalized cuts and watersheds for image segmentation. IET Conference on Image Processing. 1-6.

[Mukherjee et al. 2014] Mukherjee, R.; Manohar, D.-D.; Das, D.-K.; Achar, A.; Mitra, A.; and Chakraborty, C.; 2014. Automated

Tissue Classification Framework for Reproducible Chronic Wound Assessment. BioMed Research International 2014. 9.

[Shorten et al. 2019] Shorten, C.; Khoshgoftaar, T.-M.; 2019. A survey on Image Data Augmentation for Deep Learning. J Big Data 6, 60.

[Shenoy et al. 2018] Shenoy, V.; Foster, E.; Aalami, L.; Majeed, B.-A.; and Aalami, O.; 2018. Deepwound: Automated Postoperative Wound Assessment and Surgical Site Surveillance through Convolutional Neural Networks.

[Zhang et al. 2018] Zhang, J.; Zhu, E.; Guo, X.; Chen, H.; and Yin, J.; 2018. Chronic Wounds Image Generator Based on Deep Convolutional Generative Adversarial Network. 36th National Conference of Theoretical Computer Science. 150-158.

[Oktay et al. 2018] Oktay, O.; Schlemper, J.; Folgoc, L.-L.; Lee, M.; Heinrich, M.; Misawa, K.; Mori, K.; McDonagh, S.; Hammerla, N.-Y.; Kainz, B.; Glocker, B.; and Rueckert, D.; 2018. Attention U-Net: Learning Where to Look for the Pancreas. Medical Imaging with Deep Learning.

[Thomas 2019] Thomas, S.; 2019. Medetec Wound Database. [Online].Available:http://www.medetec.co.uk/files/medetec/imagedatabases. html

[Trusov et al. 2020] Trusov, A.; and Limonova, E.; 2020. The analysis of projective transformation algorithms for image recognition on mobile devices. Twelfth International Conference on Machine Vision. 11433.